\documentclass[final,1p,times]{elsarticle}

\usepackage{graphicx}
\usepackage{amsmath}
\usepackage{amssymb}

\usepackage[ruled,vlined]{algorithm2e}

\usepackage{hyphenat}
\usepackage[autolanguage]{numprint}

\usepackage{natbib}
\biboptions{sort&compress}

\usepackage[colorlinks]{hyperref}

\journal{Journal of Computational Physics}

\begin{document}

\begin{frontmatter}

\title{A parallel fast multipole method for elliptic difference equations}

\author{Sebastian Liska\corref{cor1}}
\ead{sliska@caltech.edu}
\author{Tim Colonius}
\ead{colonius@caltech.edu}
\cortext[cor1]{Corresponding author}
\address{Division of Engineering and Applied Science, California Institute of Technology, Pasadena, CA 91125, USA}

\begin{abstract}
A new fast multipole formulation for solving elliptic difference equations on unbounded domains and its parallel implementation are presented.
These difference equations can arise directly in the description of physical systems, e.g. crystal structures, or indirectly through the discretization of PDEs.
In the analog to solving continuous inhomogeneous differential equations using Green's functions, the proposed method uses the fundamental solution of the discrete operator on an infinite grid, or lattice Green's function.
Fast solutions $\mathcal{O}(N)$ are achieved by using a kernel-independent interpolation-based fast multipole method.
Unlike other fast multipole algorithms, our approach exploits the regularity of the underlying Cartesian grid and the efficiency of FFTs to reduce the computation time.
Our parallel implementation allows communications and computations to be overlapped and requires minimal global synchronization.
The accuracy, efficiency, and parallel performance of the method are demonstrated through numerical experiments on the discrete 3D Poisson equation.

\end{abstract}


\end{frontmatter}

\section{Introduction}
\label{sec:introduction}

Numerical simulations of physical phenomena often require fast solutions to linear, elliptic difference equations with constant coefficients on regular, unbounded domains.
These difference equations naturally arise in the description of physical phenomena including random walks \cite{mccrea1940}, crystal physics \cite{montroll1955}, and quantum mechanics \cite{economou1984}.
Additionally, such difference equations can result from the discretization of PDEs on infinite regular grids or meshes \cite{bramble1962,buneman1971,gillman2010,gillman2014}.
Apart from the accuracy with which the underlying PDE is solved, an accurate solution of the difference equations themselves is relevant for compatible spatial discretization schemes that enforce discrete conservation laws \cite{arnold2006,perot2011}.
Examples of these techniques include finite-volume methods, mimetic schemes, covolume methods, and discrete calculus methods.

The present method considers difference equations formally defined on unbounded Cartesian grids.
Solutions to the difference equations are obtained through the convolution of the fundamental solution of the discrete operator with the source terms of the difference equations. 
As a result, the formally infinite grid can be truncated to a finite computational grid by removing cells that contain negligible source strength.
The ease with which this technique is able to adapt the computational domain makes it well-suited for applications involving the temporal evolution of irregular source distributions.
For problems that are efficiently described by block-structured grids it is possible to adapt the computational domain by simply adding or removing blocks; an example of this technique applied to an incompressible flow is provided in Section~\ref{sec:conclusions}.

The fundamental solution of discrete operators on regular grids, or lattices, are often referred to as lattice Green's functions (LGFs).
Expressions for LGFs can be readily obtained in the form of Fourier integrals, but it is typically not possible to reduce the integral representations to expressions only involving a few elementary functions \cite{glasser1977,delves2001}.
The analytical treatment and the numerical evaluation of many LGFs is facilitated by the availability of asymptotic expansions \cite{duffin1958,mangad1967,martinsson2002}.
Although LGFs have been extensively studied, they have rarely been used for solving large systems of elliptic difference equations (exceptions include 2D problems \cite{martinsson2009,gillman2010,gillman2014}).
The present work extends the use of LGFs to large scale computations involving solutions to 3D elliptic difference equations.

Solving the system of difference equations using LGFs requires evaluating discrete convolutions of the form
\begin{equation}
u(\mathbf{x}_i) = [K*\mathbf{f}](\mathbf{x}_i) = \sum_{j=0}^{M-1} K(\mathbf{x}_i,\mathbf{y}_j) f(\mathbf{y}_j),\,\, i=0,1,\dots,N-1,
\label{eq:conv}
\end{equation}
where $K(\mathbf{x}_i,\mathbf{y}_j)$ is the kernel describing the influence of a source located at $\mathbf{y}_j$ with strength $f(\mathbf{y}_j)$ has on the field $u(\mathbf{x})$ at location $\mathbf{x}_i$.
For the case of $M=N$, the straightforward approach to evaluate Eq.~\ref{eq:conv} requires $\mathcal{O}(N^2)$ operations.
There are several techniques for evaluating Eq.~\ref{eq:conv} in $\mathcal{O}(N)$ or $\mathcal{O}(N \log N)$ operations.
A few of these techniques are FMMs, FFT-based methods, particle-in-cell methods, particle-mesh methods, multigrid techniques, multilevel local-correction methods, and hierarchical matrix techniques.
In the interest of brevity, a literature review of all the methods related to the fast evaluation of Eq.~\ref{eq:conv} is omitted; instead we focus our attention on FMMs.

The performance of FMMs relies on the existence of a compressed, or low-rank, representation of the far-field behavior of $K(\mathbf{x},\mathbf{y})$ that can be used to evaluate Eq.~\ref{eq:conv} to a prescribed tolerance.
Classical fast multipole methods \cite{greengard1987,greengard1997} require analytical expansions of the far-field behavior of kernels in order to derived low-rank approximations.
Although classical FMMs can be developed for the asymptotic expansion of LGFs, alternative FMMs that are better suited for complicated kernel expressions have been developed.  
Kernel-independent FMMs \cite{gimbutas2003,ying2004,martinsson2007,fong2009,zhang2011,gillman2014} do not require analytical expansions of the far-field; instead, for suitable kernels, these methods only require numerical evaluations of the kernel.

The present method is a kernel-independent interpolation-based FMM for non-oscillatory translation-invariant kernels \cite{dutt1996,fong2009}.
These FMMs achieve low-rank approximations of the kernel by projecting the kernel onto a finite basis of interpolation functions.
Interpolation-based FMMs \cite{dutt1996,fong2009} use Chebyshev interpolation and accelerate convolutions involving the compressed kernel using singular-value-decompositions (SVDs).
In contrast, our method uses polynomial interpolation on equidistant nodes and accelerates convolutions involving the compressed kernel using FFTs.
Intermediate regular grids and fast FFT-based convolutions have been used by other FMMs \cite{berman1995,phillips1997,ying2004,chatelain2010}, and have been shown to be particularly useful in accelerating the computations of 3D methods \cite{ying2004}.
The use of intermediate regular grids in our method has the added advantage of simplifying the multilevel algorithm, since sources and evaluation points are defined on Cartesian grids at all levels of the multilevel scheme.
The spatial regularity allows for the same fast convolution techniques to be used in determining near-field and far-field contributions.
In addition to the base algorithm, our method allows for pre-computations that further accelerate the solver.

The present FMM is similar to the recent 2D FMM \cite{gillman2014} in that they both solve difference equations on unbounded domains.
In contrast to our method, this method uses skeleton/proxy points and rank-revealing factorizations to obtain low-rank approximations of the kernel.
Although we think it is possible to extend this method to 3D, we refrain from speculating on the performance of the algorithm since such extensions are not explored in current literature and their details are unclear to us.

Details regarding LGFs and their relation to solving difference equations on unbounded domains are presented in Section~\ref{sec:prelim}.
This section also describes methods for performing fast convolutions based on kernel compression and FFT techniques, and presents a context in which these two techniques can be combined to yield an even faster convolution scheme.
The resulting fast multipole algorithm and its parallel extension are then described in Section~\ref{sec:flgf}.
Finally, serial and parallel numerical experiments are reported and analyzed in Section~\ref{sec:results}.

\section{Lattice Green's functions and fast block-wise convolution techniques}
\label{sec:prelim}
\subsection{Solving difference equations on infinite Cartesian grids}
\label{sec:prelim_lgf}

The method proposed in this paper is designed to solve inhomogeneous, linear, constant-coefficient difference equations on unbounded domains.
As a representative problem, we consider in detail the difference equations resulting from the discretization of Poisson's equation in 3D.
Consider the Poisson equation
\begin{equation}
[\Delta u](\mathbf{x}) = f(\mathbf{x}),\,\, supp(f) \in \Omega,
\label{eq:poisson}
\end{equation}
where $\mathbf{x} \in \mathbb{R}^3$, $\Omega$ is a bounded domain in $\mathbb{R}^3$, and $u(\mathbf{x})$ decays as $1/|\mathbf{x}|$ at infinity.
Eq.~\ref{eq:poisson} has the analytic solution
\begin{equation}
u(\mathbf{x}) = [G * f](\mathbf{x}) = \int_{\Omega} G(\mathbf{x}-\boldsymbol{\xi})f(\boldsymbol{\xi})\, d\boldsymbol{\xi} ,
\label{eq:poissonsoln}
\end{equation}
where $G(\mathbf{x}) = -1/4\pi|\mathbf{x}|$ is the fundamental solution of the Laplace operator.
Discretizing Eq.~\ref{eq:poisson} on an infinite uniform Cartesian grid using a standard second-order finite-difference or finite-volume scheme produces a set of difference equations
\begin{equation}
[\mathsf{L} \mathsf{u}]( \mathbf{n} ) = \mathsf{f}(\mathbf{n}),\,\, supp(\mathsf{f}) \in \Omega_h ,
\label{eq:dpoisson}
\end{equation}
where $\mathsf{L}$ is the standard 7-pt discrete Laplace operator, $\mathbf{n}\in\mathbb{Z}^3$, and $\Omega_h$ is a bounded domain in $\mathbb{Z}^3$.
In practice, the constraint on $supp(\mathsf{f})$ can be relaxed by prescribing a finite tolerance and requiring that all non-negligible sources, i.e. sources with a magnitude greater than the prescribed tolerance, be located in a bounded region.
The solution to Eq.~\ref{eq:dpoisson} is given by 
\begin{equation}
\mathsf{u}(\mathbf{n}) = [\mathsf{G} * \mathsf{f}](\mathbf{n}) = \sum_{\mathbf{m}\in\Omega_h} \mathsf{G}(\mathbf{n}-\mathbf{m}) \mathsf{f}(\mathbf{m}) ,
\label{eq:dpoissonsoln}
\end{equation}
where $\mathsf{G}(\mathbf{n})$ is the fundamental solution of the discrete Laplace operator.
An expression for $\mathsf{G}(\mathbf{n})$ in terms of Fourier integrals is provided by
\begin{equation}
\mathsf{G}(\mathbf{n}) = \frac{1}{8\pi^3}
\int_{[-\pi,\pi]^3}
\frac{ \exp\left(-i \mathbf{n}\cdot\boldsymbol{\xi}\right)}
{ 2\cos(\xi_1)+2\cos(\xi_2)+2\cos(\xi_3)-6 }
\,d\boldsymbol\xi .
\label{eq:lgf1}
\end{equation}
The expression in Eq.~\ref{eq:lgf1} is readily obtained by first using Discrete Fourier Transforms (DFTs) to diagonalize $\mathsf{L}$.
The diagonalized operator is then inverted and subsequently transformed back to the original space using inverse DFTs. Infinite sums in the resulting expression are converted to integrals using appropriate limiting procedures.
Details regarding the construction of Eq.~\ref{eq:lgf1} and expressions for the fundamental solutions to other discrete operators are found in \cite{mccrea1940,duffin1958,martinsson2002}.
Additionally, \ref{sec:app_lgf} provides an outline of the numerical procedures used to evaluate $\mathsf{G}(\mathbf{n})$ for small values of $|\mathbf{n}|$.

Although it is possible to compute $\mathsf{G}(\mathbf{n})$ by numerically evaluating Eq.~\ref{eq:lgf1}, for large values of $|\mathbf{n}|$ it is more efficient to evaluate the LGF via its asymptotic expansion.
Techniques for constructing asymptotic expansions of LGFs to arbitrary order are described in \cite{duffin1958,mangad1967,martinsson2002}.
Let $A^q_\mathsf{G}(\mathbf{x})$ denote the $q$-term asymptotic expansion of $\mathsf{G}(\mathbf{n})$.
For the 3D case, we define $A^q_\mathsf{G}(\mathbf{x})$ as the unique rational function such that
\begin{equation}
	\mathsf{G}(\mathbf{n}) = A^q_\mathsf{G}(\mathbf{n}) + \mathcal{O}\left( |\mathbf{n}|^{-2q-1} \right)
\end{equation}
as $|\mathbf{n}|\rightarrow \infty$.
For $q=2$, this asymptotic expansion is given by 
\begin{equation}
A^2_\mathsf{G}(\mathbf{x}) = -\frac{1}{4 \pi |\mathbf{x}|}
- \frac{x_1^4 + x_2^4 + x_3^4 - 3 x_1^2 x_2^2 - 3x_1^2 x_3^2 - 3 x_2^2 x_3^2}{16 \pi |\mathbf{x}|^7},
\label{eq:lgf2}
\end{equation}
where $\mathbf{x}=\left( x_1, x_2, x_3 \right)$.
As expected, the first term in Eq.~\ref{eq:lgf2} corresponds to the fundamental solution of the Laplace operator.
We note that, as is the case for many asymptotic expansions, it is not always possible to increase the accuracy of the expression for a fixed argument by increasing the number of terms.%
\footnote{%
For example, for $|\mathbf{n}|=10$ the minimum value of $|A^q_\mathsf{G}(\mathbf{n})-\mathsf{G}(\mathbf{n})|/\mathsf{G}(\mathbf{n})$ is approximately $10^{-7}$ and is achieved by $n=6$. Increasing or decreasing $n$, i.e. the number of terms in the asymptotic expansion, increases the relative difference between the $A^q_\mathsf{G}(\mathbf{n})$ and $\mathsf{G}(\mathbf{n})$ for $|\mathbf{n}|=10$.%
}

Despite the fact that $G(\mathbf{x})$ and $\mathsf{G}(\mathbf{n})$ share the same asymptotic behavior, there are significant differences in their behavior near the origin.
Unlike $G(\mathbf{x})$, which is singular at the origin, $\mathsf{G}(\mathbf{n})$ remains finite for all values of $\mathbf{n}$.
$G(\mathbf{x})$ is scale-invariant, i.e. there exists a $k$ such that $G(\alpha\mathbf{x}) = \alpha^{k} G(\mathbf{x})$, whereas $\mathsf{G}(\mathbf{n})$ is not scale-invariant.
Furthermore, $G(\mathbf{x})$ is spherically symmetric about the origin, as opposed to $\mathsf{G}(\mathbf{n})$, which has reflectional symmetry about the principal axes and is invariant under index permutations.

In addition to providing expressions for the fast evaluation of LGFs, asymptotic expansions of LGFs allow for the sum given in Eq.~\ref{eq:dpoissonsoln} to be decomposed into three parts:
\begin{equation}
\mathsf{u}(\mathbf{n}) = \mathsf{u}^{\text{direct}}(\mathbf{n}) + u^{\text{asympt},q}(\mathbf{n}) + \epsilon(\mathbf{n}),
\label{eq:dpsoln}
\end{equation}
where
\begin{eqnarray}
\mathsf{u}^{\text{direct}}(\mathbf{n}) &=& \sum_{\mathbf{m}\in\Omega_h^{\text{direct}}(\mathbf{n})} \mathsf{G}(\mathbf{n}-\mathbf{m}) \mathsf{f}(\mathbf{m})\,, \label{eq:dpsoln1}\\
u^{\text{asympt},q}(\mathbf{n}) &=& \sum_{\mathbf{m} \in \Omega_h \smallsetminus \Omega_h^{\text{direct}}(\mathbf{n})} A^q_\mathsf{G}(\mathbf{n}-\mathbf{m}) \mathsf{f}(\mathbf{m})\,, \label{eq:dpsoln2}
\end{eqnarray}
and $\epsilon(\mathbf{n})$ is the error due to approximating $\mathsf{G}(\mathbf{n})$ with $A^q_\mathsf{G}(\mathbf{n})$ over the region $\Omega_h \smallsetminus \Omega_h^{\text{direct}}$.
The region $\Omega_h^{\text{direct}}(\mathbf{n})$ is a subset of $\Omega_h$ for which the LGF is evaluated directly, i.e. via numerical evaluation of Eq.~\ref{eq:lgf1}, as opposed to being evaluated via its asymptotic expansion.
Typically, the region $\Omega_h^{\text{direct}}(\mathbf{n})$ is defined by a small cubic box centered at the grid point $\mathbf{n}$.%
\footnote{%
  For the case of the 3D discrete Laplace operator, choosing $\Omega_h^{\text{direct}}$ to be a cubic box with side lengths of 14, 41, and 134 grid points is sufficient to achieve relative errors less than $10^{-5}$, $10^{-10}$, and $10^{-15}$, respectively, using the five term asymptotic expansion.
  The size of $\Omega_h^{\text{direct}}$ can be reduced by including more terms in the asymptotic expansion, for example, relative errors less than $10^{-15}$ are achieved by using the thirteen term asymptotic expansion and a box with 38 grid points on each side.%
}
The first term of Eq.~\ref{eq:dpsoln}, $\mathsf{u}^{\text{direct}}(\mathbf{n})$, is a grid function evaluated at the grid point $\mathbf{n}$, whereas the second term, $u^{\text{asympt},q}(\mathbf{n})$, is a continuous function evaluated at the location of the grid point $\mathbf{n}$.
As will be discussed in subsequent sections, this decomposition allows for $u^{\text{asympt},q}(\mathbf{n})$ to be evaluated using fast techniques developed for continuous kernels.

\subsection{Fast convolutions on regular grids via FFTs}
\label{sec:prelim_convfft}

Although discrete convolutions via FFTs is a well-known technique, a brief description is provided in order to introduce procedures and notation subsequently referenced in different steps of the overall algorithm.
Consider the one-dimensional convolution given by
\begin{equation}
u(x_i) = \sum_{j=0}^{M-1} K(x_i,y_j) f(y_j)\,, i=0,1,\dots,N-1.
\label{eq:conv1d}
\end{equation}
where $x_i=x_0+ih$ for $i=0,1,\dots,N-1$, and $y_j=y_0+jh$ for $j=0,1,\dots,M-1$.
If the kernel $K(x,y)$ is translation invariant, i.e. $K(x,y)=K(x-y)$, then Eq.~\ref{eq:conv1d} can be expressed as the discrete convolution between two vectors,
\begin{equation}
{u}_i = \sum_{j=0}^{M-1} {k}_{N-1+j-i} {f}_j\,, i=0,1,\dots,N-1\,,
\label{eq:conv1dv}
\end{equation}
where ${u}_i=u(x_i)$, ${f}_j=f(x_j)$, and ${k}_{N-1+j-i}=K(x_j-y_i)$.
Discrete linear convolutions of this form can be cast into circular convolutions by appropriate padding of the vectors $u$ and $f$.
Performing these convolutions using DFTs leads to the fast FFT-based convolution technique given by
\begin{enumerate}
\item Pad sequence with zeros: append $N-1$ zeros to vector $f$.
\begin{equation}
\bar{{f}}_i=[\text{Pad}({f})]_i = \left\{
\begin{array}{ll}
{f}_i & i=0,1,\dots,N-1 \\
0 & i=N,N+1,\dots,N+M-2
\end{array}\right.
\end{equation}
\item Forward DFT: compute the DFT of sequences $\bar{{f}}$ and ${k}$ via FFTs.
\begin{equation}
\hat{f} = \text{FFT}(\bar{f}),\, \hat{{k}} = \text{FFT}({k})
\label{eq:fftconv1}
\end{equation}
\item Convolution of DFTs: multiply complex coefficients of $\hat{f}$ and $\hat{{k}}$.
\begin{equation}
\hat{{u}}_i = [\text{Prod}({g},{k})]_i = \hat{f}_i\hat{{k}}_i\,, i=0,1,\dots,N+M-2
\label{eq:fftconv2}
\end{equation}
\item Backward DFT: compute the inverse DFT of sequence $\hat{{u}}$ via FFT.
\begin{equation}
\bar{{u}} = \text{FFT}^{-1}(\hat{u})
\label{eq:fftconv3}
\end{equation}
\item Truncate sequence: remove the first $M-1$ entries of $\bar{{u}}$ to obtain ${u}$.
\begin{equation}
{u}_i = [\text{Trunc}(\bar{{u}})]_i = \bar{{u}}_{M+i}\,, i=0,1,\dots,N-1
\label{eq:fftconv4}
\end{equation}
\end{enumerate}
This technique requires $\mathcal{O}\left((N+M)\log(N+M)\right)$ operations and is readily generalized to higher dimensions for the case of tensor-product grids by recursively applying the 1D version to each directions.

\subsection{Adaptive block-structured grid}
\label{sec:prelim_grid}
Fast convolutions via FFTs discussed in Section~\ref{sec:prelim_convfft} can be used to accelerate the evaluation of Eq.~\ref{eq:dpoissonsoln}.
In order to use this technique, the support of $\mathsf{f}$ needs to be padded with zeros to form a box.
Similarly, the region where $\mathsf{u}$ is evaluated needs to be extended to also form a box.
For cases where the domain defined by the support of the source terms is not a box, the cost of the additional computational elements can outweigh the reduced operation count per grid point of the FFT-based convolution technique.

Computational domains defined by the union of blocks can, however, be used to avoid excessive padding and still retain sufficient regularity to benefit from the fast FFT-based convolution technique.
Our formulation partitions the infinite grid into blocks defined on a logically Cartesian grid.
Blocks can potentially have a different number of grid points in each direction, but all blocks are required to have the same dimensions.
An active source block denotes a block containing non-zero sources.
Similarly, an active evaluation block denotes a block containing grid points on which the induced field is evaluated.
The union of active source (evaluation) blocks is referred to as the active source (evaluation) grid.
We emphasize that grid adaptivity is achieved through the selective choice of active blocks in order to define efficient computational domains, the present method does not consider problems with multiple spatial resolutions.

Let $B_s$ and $B_e$ denote the sets of active source and evaluation blocks, respectively.
The convolution given in Eq.~\ref{eq:dpoissonsoln} can be evaluated by
\begin{equation}
{u}^{P} = \sum_{Q \in B_s} \text{conv}(k^{Q-P},f^{Q}),\,\,\forall P\in B_e ,
\label{eq:bconv}
\end{equation}
where $u^{P}$ and $f^{P}$ denote vectors containing the values of $\mathsf{u}(\mathbf{n})$ and $\mathsf{f}(\mathbf{n})$, respectively, evaluated on the grid points belonging to block $P$.
Similarly, $k^{Q-P}$ denotes the vector containing the unique values of $\mathsf{G}(\mathbf{m}-\mathbf{n})$, as described in Section~\ref{sec:prelim_convfft}, for values of $\mathbf{n}$ and $\mathbf{m}$ corresponding to the indices of grid points belonging to block $Q$ and $P$, respectively.
If blocks $P$ and $Q$ are sufficiently well-separated then $A^q_\mathsf{G}(\mathbf{m}-\mathbf{n})$ is used instead of $\mathsf{G}(\mathbf{m}-\mathbf{n})$ for constructing $k^{Q-P}$.
The operator $\text{conv}(k^{Q-P},f^{Q})$ denotes the generalization of Eq.~\ref{eq:conv1d} to arbitrary dimensions.
Details regarding the construction of vectors $u^{P}$, $f^{P}$, and $k^{Q-P}$ are omitted, since it immediately follows the discussion regarding the tensor-product grid generalization of Eq.~\ref{eq:conv1d}. 

Computing each instance of $\text{conv}(k^{Q-P},f^{Q})$ in Eq.~\ref{eq:bconv} using the fast FFT-based convolution technique leads to a scheme that evaluates Eq.~\ref{eq:dpoissonsoln}, for the case of $B_s=B_e$, in $\mathcal{O}(N_B^2 N_b \log(N_b))$ operations, where $N_B$ is the number of blocks belonging to $B_s$, and $N_b$ is the number of grid points belonging to each block.
The operation count can be further reduced to $\mathcal{O}(N_B N_b \log(N_b) + N_B^2 N_b)$ if the DFT of the kernel blocks, $\hat{k}^{Q-P}$, are pre-computed, and if the DFT of source and evaluation blocks are preformed as pre-processing and post-processing step, respectively.
Details regarding pre-computations, and pre- and post-processing steps are discussed in subsequent sections.

\subsection{Fast convolutions via interpolation-based kernel compression}
\label{sec:prelim_ifmm}

Interpolation-based FMMs obtain a low-rank representation of the kernel, $K(\mathbf{x},\mathbf{y})$, by projecting it onto a finite basis of interpolation functions.
Consider a function $f(\mathbf{x})$ sampled at $n$ points,  $\mathbf{x}_0,\mathbf{x}_1,\dots,\mathbf{x}_{n-1}$.
An approximation for $f(\mathbf{x})$ is given by
\begin{equation}
\tilde{f}^n(\mathbf{x}) = \sum_{i=0}^{n-1} \phi_i(\mathbf{x}) f(\mathbf{x}_i) ,
\label{eq:interp}
\end{equation}
where $\phi_i(\mathbf{x})$ is a interpolation function associated with the interpolation node $\mathbf{x}_i$.
An approximation for $K(\mathbf{x},\mathbf{y})$ is obtained by recursively applying Eq.~\ref{eq:interp} to each argument of $K(\mathbf{x},\mathbf{y})$,
\begin{equation}
\tilde{K}^{n}(\mathbf{x},\mathbf{y}) = 
\sum_{i=0}^{n-1} \sum_{j=0}^{n-1}
\psi_i(\mathbf{x}) K(\mathbf{x}_i,\mathbf{y}_j) \phi_j(\mathbf{y}) ,
\label{eq:approxkernel}
\end{equation}
where $\{\phi\}=\{\phi_0,\phi_1,\dots,\phi_{n-1}\}$ and $\{\psi\}=\{\psi_0,\psi_1,\dots,\psi_{m-1}\}$ are potentially distinct bases of interpolation functions.
In order to make the kernel-compression technique symmetric, only schemes with $\{\phi\}=\{\psi\}$ are considered.
Directly applying this kernel compression technique to discrete convolutions of the form of Eq.~\ref{eq:conv} leads to an approximation of $u(\mathbf{x}_i)$ given by
\begin{equation}
  u(\mathbf{x}_i) \approx
  \sum_{j=0}^{M-1}
    \sum_{p=0}^{n-1} \sum_{q=0}^{n-1}
      \phi_p(\mathbf{x}_i) K(\mathbf{x}_p,\mathbf{y}_q) \phi_q(\mathbf{y}_j) f(\mathbf{y}_j),\,\, i=0,1,\dots,N-1 .
\label{eq:ifmm}
\end{equation}
For cases involving multiple sets of either evaluation points, $\mathbf{x}_i$, or source points, $\mathbf{y}_j$, it is advantageous to decompose the evaluation of Eq.~\ref{eq:ifmm} into three steps:
\begin{enumerate}
\item Regularization: compute effective source terms using the adjoint of the interpolation procedure.
\begin{equation}
\tilde{f}(\mathbf{y}_q) =
  \sum_{j=0}^{M-1} \phi_q(\mathbf{y}_j) f(\mathbf{y}_j),\,\, q=0,1,\dots,n-1
\label{eq:kcprss1}
\end{equation}
\item Convolution: compute the field induced by effective source terms on interpolation nodes.
\begin{equation}
\tilde{u}(\mathbf{x}_p) =
  \sum_{q=0}^{n-1}
  	K(\mathbf{x}_p,\mathbf{y}_q) \tilde{f}(\mathbf{y}_q),\,\, p=0,1,\dots,n-1
\label{eq:kcprss2}
\end{equation}
\item Interpolation: compute the field at evaluation points using the interpolation procedure.
\begin{equation}
u(\mathbf{y}_i) =
  \sum_{p=0}^{n-1} \phi_p(\mathbf{x}_i) \tilde{u}(\mathbf{x}_j),\,\, i=0,1,\dots,N-1
\label{eq:kcprss3}
\end{equation}
\end{enumerate}
If the values of $\phi_q(\mathbf{x}_i)$ and $\phi_q(\mathbf{y}_j)$ are known, the number of operations required by this procedure, for the case of $M=N$, is $\mathcal{O}(2 nN + n^2)$.
For the case of $n \ll N$ this procedures represents a significant reduction in the number of operations compared to straightforward method of evaluating Eq.~\ref{eq:conv}.

\subsection{Fast convolution on regular grids using polynomial interpolation and FFTs}
\label{sec:prelim_interp}

The fast convolution techniques presented in Sections \ref{sec:prelim_convfft} and \ref{sec:prelim_ifmm} can be combined to yield a faster method for evaluating the block-wise convolutions involved in Eq.~\ref{eq:bconv}.
This technique follows from the observation that Eq.~\ref{eq:kcprss2} can be evaluated using FFTs if the kernel is translation-invariant and the interpolation nodes on which the kernel is evaluated are restricted to be on a regular grid.
The first requirement is assumed since the kernels corresponding to the fundamental solutions of the linear, constant-coefficient elliptic difference equations are translation-invariant. 
The second condition is achieved by using an interpolation scheme based on equidistant interpolation nodes on tensor-product grids.
Although many interpolation schemes satisfy the latter condition, only polynomial interpolation schemes on tensor-product grids are presently considered since they are fast, simple to implement, and their behavior is well-understood.%
\footnote{%
  Alternative interpolation procedures, with the exception of Fourier interpolation on non-periodic domains, have not been explored.
  Although Fourier interpolation is particularly appropriate given FFTs are used in our method to accelerate local computations, preliminary results have shown that this procedure is less efficient (in terms of points per unit accuracy) than the procedure described in this section.%
}

Polynomial interpolation on tensor-product grids is performed by recursively applying 1D polynomial interpolation along each direction.
This generalization has the advantage of maintaining the number of operation per grid point independent of dimension, and allows the behavior of the interpolation process to be readily generalized from its 1D version.

In the absence of rounding errors, 1D polynomial interpolants converge geometrically if the function being interpolated is analytic in a region on the complex plane near the interpolation interval.
The size and shape of this convergence region depends on the choice of interpolation nodes \cite{trefethen2000}.
The kernels being considered correspond to the asymptotic expansion of the LGFs that are only discontinuous at the origin.
Although convergence conditions should be verified for each kernel, the requirement that source and evaluation blocks be sufficiently well-separated to accurately evaluate the LGF using its asymptotic expansion is often sufficient to guarantee convergence.

Unlike Chebyshev interpolation, polynomial interpolation on equidistant nodes is ill\hyp{}conditioned.
Ill-conditioning can cause rounding errors due to finite numeric precision can be amplified.
The Lebesgue constant, $\Lambda_n(X)$, for a set of $n$ interpolation points $X={x_0,x_1,\dots,x_n-1}$, can be used to bound the growth of perturbations in the data \cite{quarteroni2010},
\begin{equation}
\max_{x \in I}| p(x) - \tilde{p}(x) | \le \Lambda_n(X) \max_{0 \le i \le n-1} | f(x_i) - \tilde{f}(x_i)|\,,
\label{eq:leb}
\end{equation}
where $I$ is the interpolation interval, $p(x)$ and $\tilde{p}(x)$ are the polynomials interpolants resulting from the nodal values $f(x_i)$ and $\tilde{f}(x_i)$, respectively.
The Lebesgue constant is given by
\begin{equation}
\Lambda(X) = \max_{x \in I} \sum_{i=0}^{n-1} |\phi_i(x)|\,,
\end{equation}
where $\phi_i(x)$ is the Lagrange characteristic polynomial associate with $x_i$.
Eq.~\ref{eq:leb} can be extended to polynomial interpolation on tensor product grids, with equal number of points and spacing in each direction, by replacing $\Lambda(X)$ with $\left(\Lambda_n(X)\right)^d$, where $d$ is the dimension of the problem.

In the limit of very large $n$, the Lebesgue constant of a set of equally spaced nodes is known to grow exponentially \cite{quarteroni2010}.
In order to avoid very large Lebesgue constants, the present scheme restricts the number of nodes used for polynomial interpolation to be at most $n_{max}$. 
If $n_{max}$ nodes are insufficient to achieve a desired interpolation error, additional nodes are added to the interval, but only the closest $n_{max}$ nodes to the evaluation point are used for interpolation.
Thus, this hybrid scheme performs both $p$- and $h$-refinement to increase the accuracy of the interpolations procedure.
As a result, geometric convergence rates are expected for $n \le n_{max}$, and polynomial convergence rates of order $n_{max}-1$ are expected for $n>n_{max}$.
The values of $n$ and $n_{max}$ required to interpolate a function $f(x)$ over an interval $I$ with an interpolation error less than $\epsilon$ are obtained in two steps:
\begin{enumerate}
\item Find the largest $n_{max}$ such that $\Lambda_{n_{max}}(X) \epsilon_{p}$ is less than $\epsilon$, where $\epsilon_p$ is the precision of the floating-point scheme.
\item Progressively increase the number of $n$ until the difference between $f(x)$ and it approximation are less than $\epsilon$.
\end{enumerate}
The same procedure can be used in higher dimensions by having $n$ and $n_{max}$ correspond to the number of interpolation points along each direction, and replacing $\Lambda_{n_{max}}(X)$ with $\left(\Lambda_{n_{max}}(X)\right)^d$.
For example, approximating an analytic function in 3D using double-precision arithmetic to relative tolerance of $\epsilon=1.25\times10^{-12}$ requires $n_{max} \le 10$.

We omit a step-wise description of the combined fast algorithm for block-wise convolutions, since it readily follows from the discussion.
Instead, we introduce the notation $f^P = \text{Interp}(f^Q)$ and $f^P = \text{Reg}(f^Q)$ to denote the interpolation and regularization (adjoint of interpolation) operations, respectively, 
Instead, we introduce use the notation $f^P = \text{Interp}(f^Q)$ and $f^P = \text{Reg}(f^Q)$, respectively, to denote the interpolation and regularization (adjoint of interpolation) of $\mathsf{f}$ from block/interval $Q$ to block/interval $P$.

\section{The Fast Lattice Green's Function method}
\label{sec:flgf}

\subsection{Basic algorithm}
\label{sec:flgf_basic}

Thus far we have discussed methods for accelerating the evaluation of Eq.~\ref{eq:bconv} by performing fast block-wise convolutions involving interpolation-based kernel compression and/or FFT techniques.
Asymptotically these schemes require $\mathcal{O}(N^2)$ operations, though the constant in front of the $N^2$ term can be significantly smaller compared to that of the straightforward method.
For kernels that decay or exhibit progressively smoother behavior away from the origin, e.g. the fundamental solution of the discrete Laplace operator, it is possible to combine the fast block-wise convolution techniques discussed in Sections~\ref{sec:prelim} with the multilevel scheme of the original FMM \cite{greengard1987}.
To facilitate the discussion, we will assume that the active evaluation grid is the same as the active source grid.

Our multilevel scheme follows a tree (octree in 3D) structure similar to that described in \cite{greengard1987}.
Tree nodes at all levels are said to correspond to intervals.%
\footnote{%
  In the context of the hierarchical algorithm and structure, the term ``interval'' is equivalent to term ``box'' used in \cite{greengard1987}.
  We reserve the term ``box'' for geometric descriptions, and do not associate any specific structure or information with the term.%
}
Each interval is defined by the tensor-product grid associated with the nodes of the interpolation scheme.
The tree is constructed by first creating one tree leaf, i.e. tree node with no children, for each active grid block.
The intervals of tree leaves are defined such that they occupy the same spatial region as their associated grid blocks.
After defining all tree leaves, siblings are recursively merged to generate the multilevel structure.
We use the convention that all tree leaves are located at level $1$ and that the root of the tree is located at level $L$.

The set of intervals at level $\ell$ is denoted by $B^\ell$, and $N_B^\ell$ denotes the size of $B^\ell$.
In order to facilitate the discussion, intervals at level $\ell$ are chosen to contain $n^\ell_b$ nodes in each directions.
The total number of points in each interval at level $\ell$ is given by $N_b^\ell=(n^\ell_b)^d$, where $d$ is the dimension of the problem.
The set of blocks defining the underlying active grid is denoted by $B^0$.
$N_B^0$, $n^0_b$, and $N_b^0$ have definitions analogous to $N_B^\ell$, $n^\ell_b$, and $N_b^\ell$, respectively.
We note that level zero, $\ell=0$, is not part of the tree structure, but the slight abuse of notation facilitates the description of the algorithm.

By construction, all intervals are Cartesian grids.
As a result the union of intervals belonging to the same level defines an analogous grid to that of the underlying adaptive block-structured grid.
Therefore, the techniques for fast block-wise convolutions discussed in Section~\ref{sec:prelim} are readily generalized to all levels of the tree structure.

Our overall algorithm for solving systems of difference equations of the form given by Eq.~\ref{eq:dpoisson} on adaptive block-structured grids is referred to as the Fast Lattice Green's Function (FLGF) method.
The FLGF method is described in the following steps:
\begin{enumerate}
\setcounter{enumi}{-1}
\item \emph{Pre-computation}: compute and store all unique $\hat{k}^{P-Q}$ used in Step~2.
	\begin{equation}
		\hat{k}^{P} = \text{FFT}( {k}^{P-Q} )
		\label{eq:precomp}
	\end{equation}
\item \emph{Upwards Pass}: $\forall P \in B^\ell$, for $\ell=0,1,\dots,L$
	\begin{enumerate}
		\item Regularize: compute effective source terms at interpolation nodes
		\begin{equation}
			\tilde{f}^{P} = \sum_{Q\in\text{RegSupp}(P)} \text{Reg}( \tilde{f}^{Q} ),
			\label{eq:alg_reg}
		\end{equation}
		\item Padded forward DFT: prepare vectors for DFT convolutions
		\begin{equation}
			\hat{f}^{P} = \text{FFT}( \text{Pad}(\tilde{f}^{P} )),
			\label{eq:alg_upfft}
		\end{equation}
	\end{enumerate}
\item \emph{Level Interactions}: $\forall P \in B^\ell$, for $\ell=0,1,\dots,L$
	\begin{equation}
		\hat{u}^{P} = \sum_{Q\in\text{InflList}(P)}\text{Prod}(\hat{f}^{Q},\hat{k}^{P-Q}),
		\label{eq:alg_mult}
	\end{equation}	
\item \emph{Downwards Pass}: $\forall P \in B^\ell$, for $\ell=L,L-1,\dots,0$
	\begin{enumerate}
		\item Truncated backwards DFT: extract relevant data from DFT convolution
		\begin{equation}
			\tilde{v}^{P} = \text{Trunc}( \text{FFT}^{-1}(\hat{u}^{P}) )
			\label{eq:alg_dwfft}
		\end{equation}
		\item Interpolate: compute and aggregate the induced field at interpolation nodes
		\begin{equation}
			\tilde{u}^{P} = \tilde{v}^{P} + \text{Interp}(\tilde{u}^{\text{IntrpSupp}(P)})
			\label{eq:alg_interp}
		\end{equation}
	\end{enumerate}
\end{enumerate}
We note that the operations performed in Steps 1, 2, and 3 are commonly referred to as the \emph{multipole\hyp{}to\hyp{}multipole}, \emph{multipole\hyp{}to\hyp{}local}, and \emph{local\hyp{}to\hyp{}local} operations, respectively, in the FMM literature.
The lists of blocks/intervals used by the algorithm are given by
\begin{gather}
	\text{RegSupp}(P) = \left\{
    \begin{array}{ll}
      \text{block } P & \text{if } P \in B^{0} \\
      \text{block}\,\boldsymbol{\sim}\,\text{interval } P & \text{if } P \in B^{1} \\
      \text{children of interval } P & \text{if } P \in B^{\ell},\,2 \le \ell \le L
    \end{array}
    \right. \\
    \text{IntrpSupp}(P) = \left\{
      \begin{array}{ll}
        \text{interval}\,\boldsymbol{\sim}\,\text{block } P & \text{if } P \in B^{0} \\
        \text{parent of interval } P & \text{if } P \in B^{\ell},\,1 \le \ell < L \\
        \emptyset & \text{if } P \in B^{L}
      \end{array}
    \right. \\
    \text{InflList}(P) = \left\{
      \begin{array}{ll}
        \text{near-neighbors of block } P & \text{if } P \in B^{0} \\
        \text{interaction list of interval } P & \text{if } P \in B^{\ell},\,1 \le \ell \le L
      \end{array}
	\right.
\end{gather}
where the symbol $\boldsymbol{\sim}$ is taken here to mean \emph{associated with}.
Children, parents, near-neighbors, and interaction lists follow the same definitions those of the original FMM \cite{greengard1987}.
Based on these definitions, we note that Eq.~\ref{eq:alg_reg} reduces to $\tilde{f}^{P} = {f}^{P}$ for $P \in B^0$, and Eq.~\ref{eq:alg_interp} reduces to $\tilde{u}^{P} = \tilde{v}^{P}$ for $P \in B^L$.

\subsection{Algorithmic complexity}
\label{sec:flgf_complexity}

The overall complexity of our algorithm is $\mathcal{O}(N)$, as is the case for the original FMM.
For simplicity, the discussion concerning the cost of each step is limited to the 3D version of the algorithm.
Details regarding to the cost of each block/interval are presented in Table~\ref{tab:cost}.
The factor of $8$ in front of $C^\ell_\text{Interp}$ for the \emph{Upwards Pass} ($\ell>1$) is due to the fact that each interval has eight children.
The constants, $27$ and $189$, associated with the cost of \emph{Level Interactions} correspond to the number of near-neighbors and the number of members of each interaction list, respectively.

\begin{table}[htbp]
\centering
\caption[%
  Operation counts for each step of the LGF-FMM.
 ]{%
  Operation counts per interval/block for each step of the FLGF method.\label{tab:cost}%
}
\begin{tabular}{|lll|}
\hline
 & cost & order \\ \hline
Pre-computations & $C^\ell_\text{EvalKernel} + C^\ell_\text{PadFFT}$ & $N^\ell_b \log N^\ell_b$ \\
Upwards pass ($\ell=0$) & $C^0_\text{PadFFT}$ & $N^0_b \log N^0_b$ \\
Upwards pass ($\ell=1$) & $C^\ell_\text{Interp} + C^\ell_\text{PadFFT}$ & $N^\ell_b \log N^\ell_b$ \\
Upwards pass ($\ell>1$) & $8C^\ell_\text{Interp} + C^\ell_\text{PadFFT}$ & $N^\ell_b \log N^\ell_b$ \\
Level interactions ($\ell=0$) & $27 C^\ell_\text{Prod}$ & $27 N^0_b$ \\
Level interactions ($\ell>0$) & $189 C^\ell_\text{Prod}$ & $189 N^\ell_b$ \\
Downwards pass & $C^\ell_\text{Interp} + C^\ell_\text{PadFFT}$ & $N^\ell_b \log N^\ell_b$ \\ \hline
\end{tabular}
\end{table}

The specific values and a brief discussion of the constants presented in Table~\ref{tab:cost} are provided:
\begin{enumerate}
  \item $C^\ell_\text{EvalKernel}$: Cost of kernel evaluation performed in Eq.~\ref{eq:precomp}.
  Constructing the vector $k^{Q-P}$, where $Q$ and $P$ are blocks/intervals at level $\ell$, requires $8N^\ell_b$ kernel evaluations. 
  For small values of $|\mathbf{n}-\mathbf{m}|$ a look-up table is used to evaluate $\mathsf{G}(\mathbf{n}-\mathbf{m})$; otherwise the kernel is evaluated using $A^n_\mathsf{G}(\mathbf{n}-\mathbf{m})$.
\item $C^\ell_\text{Interp}$: Cost of polynomial interpolation performed in Eq.~\ref{eq:alg_interp}.
  The coefficient mapping interpolation nodes to evaluation nodes are precomputed (only needed for 1D interpolation); therefore computing the values of a single block/interval at level $\ell$ requires $N^\ell_b \times \min\left(n^{\ell+1}_b,n_\text{max}\right)$ operations (1 real addition and 1 real multiplication per operation), where $n_\text{max}$ is described in Section~\ref{sec:prelim_interp}.
  In 3D, $n_\text{max}$ is typically set to be no greater than $10$.
$C^\ell_\text{Interp}$ also describes the cost of performing the regularization, adjoint of interpolation, operation involved in each term of the sum in Eq.~\ref{eq:alg_reg}.
  \item $C^\ell_\text{PadFFT}$: Cost of performing a 3D FFT (real-to-complex) or inverse FFT (complex-to-real) on the padded vectors present in Eq.~\ref{eq:precomp}, \ref{eq:alg_upfft}, and \ref{eq:alg_dwfft}.
  The operation count (total number of real additions and multiplications) for each FFT performed using the FFTW library is approximately $2(8N^\ell_b)\log_2(8N^\ell_b)$ \cite{johnson2007}, where $8N^\ell_b$ is the size of the padded vectors.
  Since all FFTs are real-to-complex or complex-to-real approximately half of the coefficients are redundant and neither need be stored nor operated on.
  \item $C^\ell_\text{Prod}$: Cost of performing DFT convolutions, i.e. multiplication of complex coefficients, in Eq.~\ref{eq:alg_mult}.
  If blocks/intervals $Q$ and $P$ belong to level $\ell$, then performing $\text{Prod}(\hat{f}^{P},\hat{k}^{P-Q})$ requires approximately $4N^\ell_b$ operations (1 complex multiplication per operation), where $4N^\ell_b$ is the number of non-redundant DFT coefficients per block/interval.
\end{enumerate}

The cost per level of each operation, except for the \emph{Pre-computation} step, can be obtained by multiplying the values of each row by the $N^\ell_B$.
In regards to the \emph{Pre-computation} step, the cost per level for $\ell=0$ and $\ell>0$ is obtained by multiplying the cost per block/interval by $27$ and $317$, respectively, which correspond to the number of unique $\hat{k}^{P-Q}$ vectors that are used at each level.
If the kernel shares the same symmetry as the LGF of the discrete Laplace operator, the number of unique $\hat{k}^{P-Q}$ vectors per level is reduced to $4$ and $36$ for $\ell=0$ and $\ell>0$, respectively.
If symmetry is used to reduce number of pre-computed $\hat{k}^{P-Q}$ vectors, then $C^\ell_\text{Prod}$ is roughly doubled since a twiddle factor needs to be applied to the DFT coefficients for cases involving reflections.

\subsection{Parallel implementation}
\label{sec:flgf_parallel}
A brief overview of our MPI-based algorithm is included to demonstrate that the present method allows for a simple parallel implementation suitable for practical large-scale scientific computing.
In the present implementation the tree structure and load balancing estimates are redundantly computed (in serial) by all MPI-processes.%
\footnote{%
Equivalently, the tree structure and load balancing estimates could be computed by a single MPI-process and then scattered to all processes, but this approach would result in additional communication costs.}
As a result, all the information necessary to evaluate $\text{RegSupp}$, $\text{IntrpSupp}$, and $\text{InflList}$ for any block/interval is known by all processes.
The tree structure is constructed following the bottom-up approach discussed in Section~\ref{sec:flgf_basic}.
Prior to partitioning the problem, the load balancing scheme first assigns a weight to every block and interval based on an estimate of its runtime cost.
Next, parent tree nodes are recursively grouped with one of its child tree nodes, and tree leaves are grouped with their associated grid blocks.
The set of groups is then partitioned into clusters in such a way that the weight of each cluster (aggregate weight of all interval/blocks belonging to all groups in the cluster) is roughly the same.
Each cluster is then assigned to an MPI process.
Given each group contains only one block, a Morton or Z-order curve \cite{morton1966} is used to preserve data locality during partitioning.

The parallel algorithm closely resembles the serial algorithm, since each process executes steps analogous to the \emph{Upwards Pass}, \emph{Level Interactions}, and \emph{Downwards Pass}.
Non-blocking routines are used for all communications, allowing for computations to be overlapped with communications.
Furthermore, our algorithms use these routines to avoid any global synchronization within each steps and between steps.

The parallel execution of each step follows a similar event-driven paradigm, where processes perform particular ``work units'' based on the information that has been received or is locally available, and send information to other processes as soon as a set of ``work units'' have been completed.
Send and receive buffers are used to avoid excessive memory requirements.
The algorithm gives priority to ``work units'' yielding results that are sent to other processes.
The time spent waiting to either receive information or to clear send buffers is used to perform local ``work units'', i.e. operations that only require data and yield results pertinent to the same process.
During the \emph{Upward} and \emph{Downward Pass} a non-blocking send is posted after each interval-/block-wise regulation and interpolation operation is completed.
In contrast, during \emph{Level Interactions} step the influence of all intervals/blocks belonging to a process on all intervals/blocks belonging another process is aggregated and packaged before sending; thus each process sends at most one message to every other process during this step.
For convenience, pseudo-codes for the communication patterns described in this section are given in \ref{sec:app_parallel}.

\section{Numerical results}
\label{sec:results}

In this section we present numerical results that demonstrate the accuracy, computational cost, and parallel performance of the FLGF method.
The results reported are for the discrete Poisson equation in 3D.
We clarify that the solution error is measured with respect to the exact solution of the \emph{discrete} Poisson problem, as opposed to the exact solution of the \emph{continuous} Poisson problem.

As in Section~\ref{sec:flgf}, the active evaluation grid is defined to be equal to the active source grid.
For all test cases, blocks/intervals belonging to level $\ell$ are chosen to contain $n^\ell_b$ points in each directions.
In the interest of brevity, we only consider schemes where the number of interpolation nodes per interval is the same for all levels, $n^{\ell}_b=n_I$ for $\ell=1,2,\dots,L$, and require that $n_I=n^0_b+1$.%
\footnote{%
  Our implementation requires that intervals belonging to $\ell>0$ have a one grid point overlap with their neighboring intervals along $(n_1,n_2,n_3)$ directions, where $n_i=\{0,1\}$ and at least one $n_i$ is non-zero.%
}
These considerations reduce the parameter space of possible schemes to cases where the spacing between nodes of parent intervals is twice that of child intervals.
Furthermore, there is effectively no regularization/interpolation between level 0 and level 1, since the nodes on level 1 coincide with the underlying grid points. 
Following the discussion of Section~\ref{sec:prelim_interp} we set the maximum number of nodes used for interpolating a value at single point, $n_\text{max}$, to be $10$ for all test cases.

We note that the previous restrictions are only imposed for the purpose of a concise exposition.
The non-scale-invariant behavior of most LGFs suggests that, in general, non-trivial performance gains can be achieved by tuning $n^{\ell}_b$ for each level.
Given that the LGF of the discrete Laplace operator is approximately scale-invariant away from the origin, possible performance gains achieved by varying $n^{\ell}_b$ are not considered in the present discussion.

The present MPI-based implementation is written in Fortran and makes use of the FFTW3 \cite{johnson2007} library to compute FFTs.
Numerical experiments are performed using our local computing facility which consists of 60 compute nodes connected by a QDR InfiniBand network.
Each node contains 2 Intel Xeon X5650 processors (6-core, 12MB cache, 2.66GHz clock speed) and 48GB of RAM.

\subsection{Error}
\label{sec:results_error}

\begin{figure}[htbp]
\centering
\includegraphics[width=1.0\textwidth]{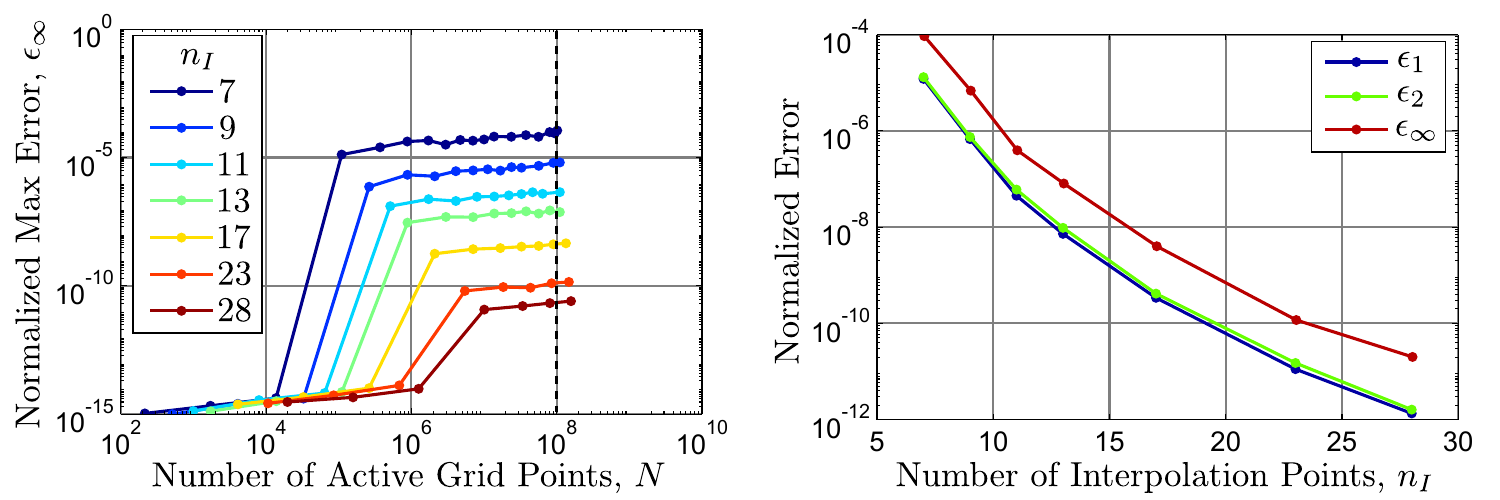}
\caption[%
  Computed interpolation error of selected LGF-FMM schemes for different problem sizes.
]{%
  \emph{Left}: max error, $\epsilon_\infty$, for cubic active grids containing $N$ grid points partitioned into blocks with $n^0_b=n_I-1$ grid points along each direction.
  \emph{Right}: error for test cases containing $10^8$ grid points as function of $n_I$; the curve corresponding to $\epsilon_\infty$ is equivalent to the values on the \emph{left} plot intersecting the vertical dashed line.\label{fig:error}%
}
\end{figure}

The accuracy of the proposed methodology is investigated on cubic active grids containing different numbers of active grid points and partitioned into blocks of different sizes.
A procedure based on random manufactured solutions is used to determine the error of each test case.
In this procedure a solution, $\mathsf{u}_\text{rand}(\mathbf{n})$, is manufactured by assigning a random value between $-1$ and $+1$ to each grid points, except for grid points on the boundary of the active grid which are set to zero.
The source distribution, $\mathsf{f}(\mathbf{n})$, which serves as the input for each test case, is computed by taking the discrete Laplacian of the prescribed solution, i.e. $\mathsf{f}(\mathbf{n})=[\mathsf{L}\mathsf{u}_\text{rand}](\mathbf{n})$.
The normalized error for each test case is computed by $\epsilon_p = ||\mathsf{u} - \mathsf{u}_\text{rand}||_p/||\mathsf{u}_p||_p$, where $\mathsf{u}(\mathbf{n})$ is determined by solving $[\mathsf{L}\mathsf{u}_p](\mathbf{n}) = \mathsf{f}(\mathbf{n})$, and $||\mathsf{u}||_p$ is the $L^p$-norm of $\mathsf{u}$ computed over the active grid.
The error for various problem sizes and schemes based on different blocks sizes is presented in Figure~\ref{fig:error}.

Test cases involving a small number blocks do not make use of the interpolation-based kernel compression technique; they only make use of the block-wise FFT-based convolution technique, which incurs an error close to machine precision.
As a result, the three cases with the smallest values of $N$ for each series have significantly smaller errors compared to their respective asymptotic (large $N$) error.

Given that we chose $n_\text{max}=10$ and constrained $n^0_b$ to be proportional to $n_I$, the error is expected to decay as $n_I^{-10}$ for $n_I>n_\text{max}$ (one order greater than interpolation order due to the $\sim|\mathbf{x}|^{-1}$ decay of the LGF).
The data shown in Figure~\ref{fig:error} are consistent with our estimates, exhibiting a behavior proportional to $n_I^{-10.7}$ for values of $n_I$ between $11$ and $28$.
The significant difference in the magnitude of $\epsilon_1$ and $\epsilon_2$ compared to $\epsilon_\infty$ suggests that the maximum error is concentrated in lower dimensional regions of the grid.
Spatial plots of the error (not included) confirm that larger errors are always observed near or on the boundary points of blocks.
These observations are characteristic of the interpolation scheme being used, which is known to exhibit larger errors near the boundaries of the interpolation interval (Runge phenomena).

Although the error for schemes with $n_I \ne n^0_b+1$ is not presented, it is readily deduced from our reported results that for any choice of $n^0_b$ the error can be controlled by changing $n_I$.
Furthermore, different choices of $n_\text{max}$ have not been explored since $n_\text{max}=10$ allows for schemes with errors as small as $\sim10^{-12}$, which are sufficient for many practical applications.
Errors smaller than $10^{-12}$ can be obtained by reducing $n_\text{max}$ and increasing $n_I$.

\subsection{Computation time}
\label{sec:results_time}
\begin{figure}[htbp]
\centering
\includegraphics[width=1.00\textwidth]{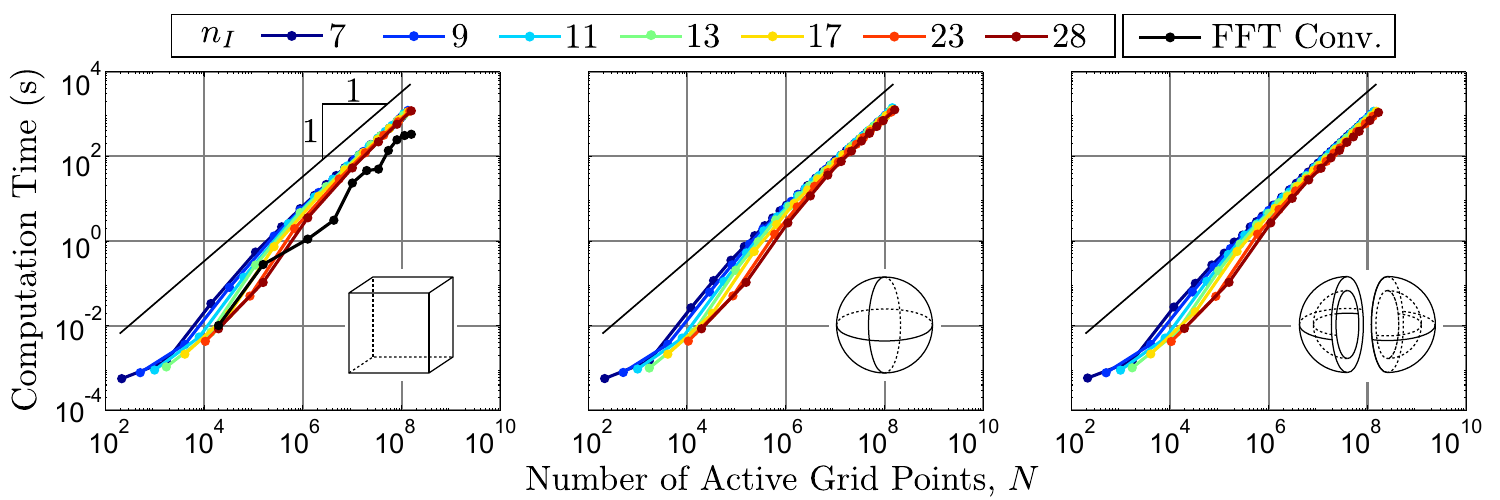}
\caption[%
  Computation times of the LGF-FMM for different source distributions and problems sizes.
]{%
    Computation times for active grids containing $N$ grid points partitioned into blocks with $n^0_b=n_I-1$ point in each direction.
    The curve labeled ``FFT Conv.'' corresponds to the special cases where the entire active grid is a single block.
    Results are presented for active grids with geometries approximating: cubes (\emph{left}), spheres (\emph{middle}), and spherical-shells (\emph{right}).\label{fig:runtime}%
}
\end{figure}

The test cases used to examine the computation time of the FLGF method follow the same setup as in Section~\ref{sec:results_error}, except that we now consider three types of active grid geometries: cubes, spheres, and spherical-shells (with a thickness of $0.1$ diameters).
For cases of spheres and spherical-shells, the set of blocks that defines the active grid is constructed so as to approximate these geometries.
Computation times for various schemes and problem sizes are presented in Figure~\ref{fig:runtime}, and asymptotic computation rates (number of active grid points per computation time) for a few schemes are included in Table~\ref{tab:rates}.

\begin{table}[htbp]
\centering
\caption[%
  Computation rates of the LGF-FMM for selected test cases.
 ]{%
  Approximate asymptotic computation rates for selected test cases presented in Figure~\ref{fig:runtime}. Values given in units of $10^5$ pts/s.
  Rates are based on the test cases with $10^8$ active grid points (values interpolated from nearest two data points).
  Asterisk $(*)$ indicates rates that are not strictly asymptotic since they correspond to $\mathcal{O}(N\log N)$ schemes.\label{tab:rates}%
}
\begin{tabular*}{0.75\textwidth}{@{\extracolsep{\fill}}|c|ccc|}
\hline
 scheme & {box} & {sphere} & {spherical-shell} \\ \hline
 $n_I =  7$ & $1.187$ & $1.167$ & $1.267$ \\
 $n_I = 13$ & $1.189$ & $1.169$ & $1.315$ \\
 $n_I = 28$ & $1.384$ & $1.341$ & $1.150$ \\
 FFT Conv. & $3.540^*$ & n/a & n/a \\ \hline
\end{tabular*}
\end{table}

As expected, the results for all schemes presented in Figure~\ref{fig:runtime}, with the exception of the one labeled ``FFT Conv.'', have an asymptotic computational complexity of $\mathcal{O}(N)$.
FFT Conv. refers to the special case where the entire active grid is a single block for which the FLGF method reduces to a single FFT-based convolution.
We note that in 3D the operation count of an FFT Conv. is roughly 8 times the operation count of solving the systems of equations obtained from the spectral discretization of differential operators on periodic domains using FFTs (referred to as ``FFT Periodic'' in the subsequent discussion).
FFT Conv. is a useful point of comparison since (1) many methods can readily consider the case of a single block with uniformly distributed sources, (2) it is well-established that for regular grids FFT-based elliptic solvers achieve very high, if not the highest, computation rates, and (3) the performance of an algorithm relative to FFT Conv. is approximately hardware independent.

Figure~\ref{fig:runtime} demonstrates that the computation rates of our multi-block algorithm are within a factor of $10$ of those corresponding to FFT Conv..
For large problems (our interest here), e.g. $N=\mathcal{O}(10^8)$, Table~\ref{tab:rates} indicates that our algorithm achieves computation rates that are roughly a third of the rate of FFT Conv. with up to 10 digits of accuracy.
By comparison, the next closest method to ours, the 2D LGF FMM of Gillman and Martinsson \cite{gillman2014}, is claimed to be two orders of magnitude slower than an FFT Periodic.
Even after accounting for the fact that in 2D an FFT Conv. is a factor of 4 slower than an FFT Periodic, we observe that our method has a significantly higher computation rate.%
\footnote{%
  For the numerical experiments reported, \cite{gillman2014} stated that ``the method was run at a requested relative precision of $10^{-10}$''.
  Given that \cite{gillman2014} did not report the error of the experiments, we assume that computation rates quoted are for approximately 10 digits of accuracy, which is the same accuracy of the rates reported for the present method.
}
We also compare the performance of the present method to that of the black-box FMM of Fong and Darve \cite{fong2009}.
In solving a 3D free-space Poisson problem with $10^6$ uniformly distributed point sources over a cube with strengths $\pm 1$, \cite{fong2009} reported a computation rate of about $1.8 \times 10^3$ points per second for 8 digits of accuracy. %
\footnote{%
  We note that 8 digits is the maximum accuracy \cite{fong2009} reported for the 3D Laplace kernel. Additionally, we note that the hardware \cite{fong2009} used to perform the numerical experiments is not reported; therefore no attempt has been made to account for hardware differences.%
}
Our method achieves a rate of about $1.2 \times 10^5$ points per second for 10 digits of accuracy, a speedup of about 65 times even with 2 more digits of accuracy.
Finally, we observe that our computation rates are comparable with those of the adaptive (locally-refined) volume FMM of Langston et al. \cite{langston2011}.
In solving a 3D free-space Poisson problem in a cube with sources corresponding to Gaussian bump solution, \cite{langston2011} reported computation rates between $2.6 \times 10^{4}$ and $1.3 \times 10^{5}$ points per second for cases with either $7$ or $8$ digits of accuracy (with number of unknowns between $2.8 \times 10^6$ and $2.6 \times 10^7$).%
\footnote{%
  The numerical experiments reported in \cite{langston2011} were performed using a shared-memory (OpenMP) implementation running on 16 cores of an Intel Xeon X7560 (2.27 GHz) based system.
  The rates included in the main text correspond to the rates we would expect to observe if the numerical experiments of \cite{langston2011} were performed on a single core of our local system; both the parallel efficiency (reported to be approximately 75\% for 16 cores) and the difference in processor clock speed have been accounted for.}

A detail report of the computation time of the \emph{Pre-computation} step is omitted, since this step is observed to require only a small fraction of the computation time of a single solve.
We substantiate this claim by reporting that for test cases involving cubic active grids containing $\mathcal{O}(10^1)$, $\mathcal{O}(10^2)$, and $\mathcal{O}(10^4)$ blocks the \emph{Pre-computation} step required less than $10\%$, $1\%$, and $0.1\%$, respectively, of the computation time of a single solve.
These results are consistent with the fact that the operation count of the pre-computation step increases with the number of levels, which in turn increase logarithmically with the number of blocks for test cases involving cubic active grids. 

\subsection{Parallel performance}
\label{sec:parallel_tests}
\begin{figure}[htbp]
\centering
\includegraphics[width=1.00\textwidth]{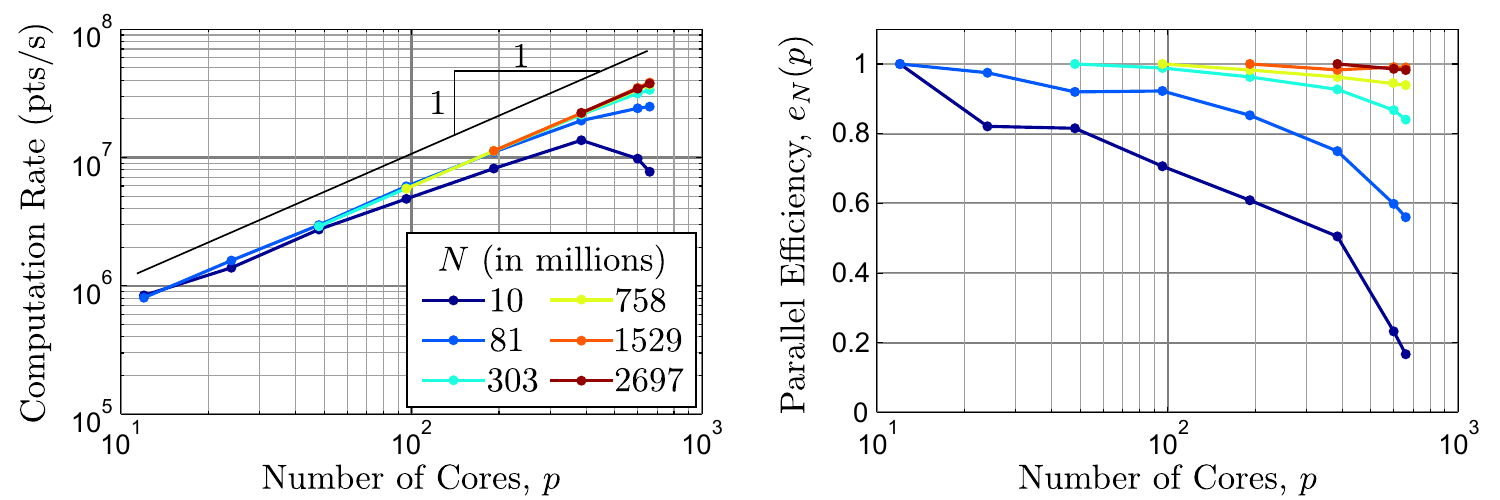}
\caption[%
  Computation rate and parallel efficiency of the LGF-FMM for different problem sizes and processor counts.
]{%
  Computation rates (\emph{left}) and parallel efficiencies (\emph{right}) for cubic active grid of various sizes.
  The parameter $n_I$ is set to $13$ for all test cases.
  The listed values of $N$, in ascending order, correspond to problems containing $5.8\times10^3$, $4.6\times10^4$, $1.8\times10^5$, $4.4\times10^5$, $8.3\times10^5$, and $1.6\times10^6$ active blocks.\label{fig:parallel}%
}
\end{figure}

The parallel performance of the FLGF method is investigated by considering cubic active grids and using the scheme corresponding to $n_I=13$ described in Sections~\ref{sec:results_error} and \ref{sec:results_time}.
Computation rates and parallel efficiencies for various problem sizes with core counts between $12$ and $660$ are included in Figure~\ref{fig:parallel}.
For all reported test cases the number of cores is a multiple of $12$ (there are $12$ cores per node in our test machine), and each MPI-process is mapped to a single core.
The parallel efficiency for each test series is defined by
\begin{equation}
e_N(p) = \frac{p_\text{min}}{p} \frac{T_N(p_\text{min})}{T_N(p)},
\label{eq:peff}
\end{equation}
where $N$ is the total number of active grid points in the test series, $p$ is the number of cores, $p_\text{min}$ is the minimal number of cores considered in the test series, and $T_N(p)$ is the runtime of the test problem.

Both strong and weak scaling can be inferred from the left plot in Figure~\ref{fig:parallel}.
Strong scaling, i.e. fixed $N$ and increasing $p$, corresponds to the individual curves associated with each test series.
Weak scaling, i.e. fixed $N/p$ and increasing $p$, is achieved when the curves associated with different test series collapse.
Figure~\ref{fig:parallel} demonstrates that, over a reasonable range $p$, the curves for most of the test series collapse to a single line with a slope approximately equal to unity.
This indicates that our implementation exhibits both good strong and weak scaling.

There are two main considerations that affect the performance of our parallel implementation.
The first is the number of blocks per core.
The FLGF method is broken into block-wise operations.
If there are too few blocks per core our total work cannot be evenly distributed across all cores.
Furthermore, given our communication scheme for the \emph{Level Interactions} step, fewer blocks per core are likely to increase the total number of MPI messages sent and received.
The second consideration is the amount work per core.
If the work per core is too small then communication cost can take an overwhelming fraction of the net run-time.
Based on these considerations and the reported results, we conclude that if each core has, on average, more than 300,000 active grid points and 200 blocks, then the parallel efficiency, as defined in Eq.~\ref{eq:peff}, is expected to be above $80\%$.
This observation seems consistent with the results reported on other MPI-based implementations of kernel-independent FMMs, for example \cite{ying2003,gholaminejad2014}.

In the interest of completeness, we note that computation rates reported in Figure~\ref{fig:parallel}, in particular those corresponding to $12$ cores, are roughly half the rates expected based on the our serial results.
This decrease in performance is due to an increase in cache-misses when more than one core per node is used.
We expect that future higher-quality implementations of the FLGF method can readily mitigate this feature.

\section{Conclusions}
\label{sec:conclusions}

We have presented a new kernel-independent fast multipole method for elliptic difference equations on infinite Cartesian grids.
The FLGF method exploits the regularity of the underlying grid to achieve small computation times by using a fast convolution technique that combines interpolation-based kernel compression and FFTs.
Interpolation based on equidistant nodes, along with a $p$- and $h$-refinement technique, is shown to be an effective scheme for obtaining low-rank representations of kernels, while still preserving sufficient regularity to allow discrete convolutions to be performed quickly using FFTs.
The adaptive block-structured grid strategy blends well with the overall algorithm, and the reported numerical experiments demonstrate that computation rates remain roughly invariant of source distributions.

The efficiency of the FLGF method is demonstrated through several numerical experiments solving the discrete 3D Poisson equation for cases involving up to 2 billion grid points and 660 cores.
Serial test cases confirm that the algorithm archives an asymptotic linear complexity. 
Computation rates of approximately $1.2\times10^5\,\text{pts/s}$ or, equivalently, grind times of $8.3\mu\text{s}$ are observed for problems containing $10^8$ grid points.
The computation rate is shown to be roughly invariant to different source distributions and block sizes.
Furthermore, the time required to perform all pre-computations for typical problems is shown to negligible.
Test cases investigating the parallel performance of our implementation demonstrate that parallel efficiencies higher than $80\%$ are achieved under modest considerations (at least $200$ blocks and $300,000$ grid points per core).

\begin{figure}[htbp]
\centering
\includegraphics[width=0.75\textwidth]{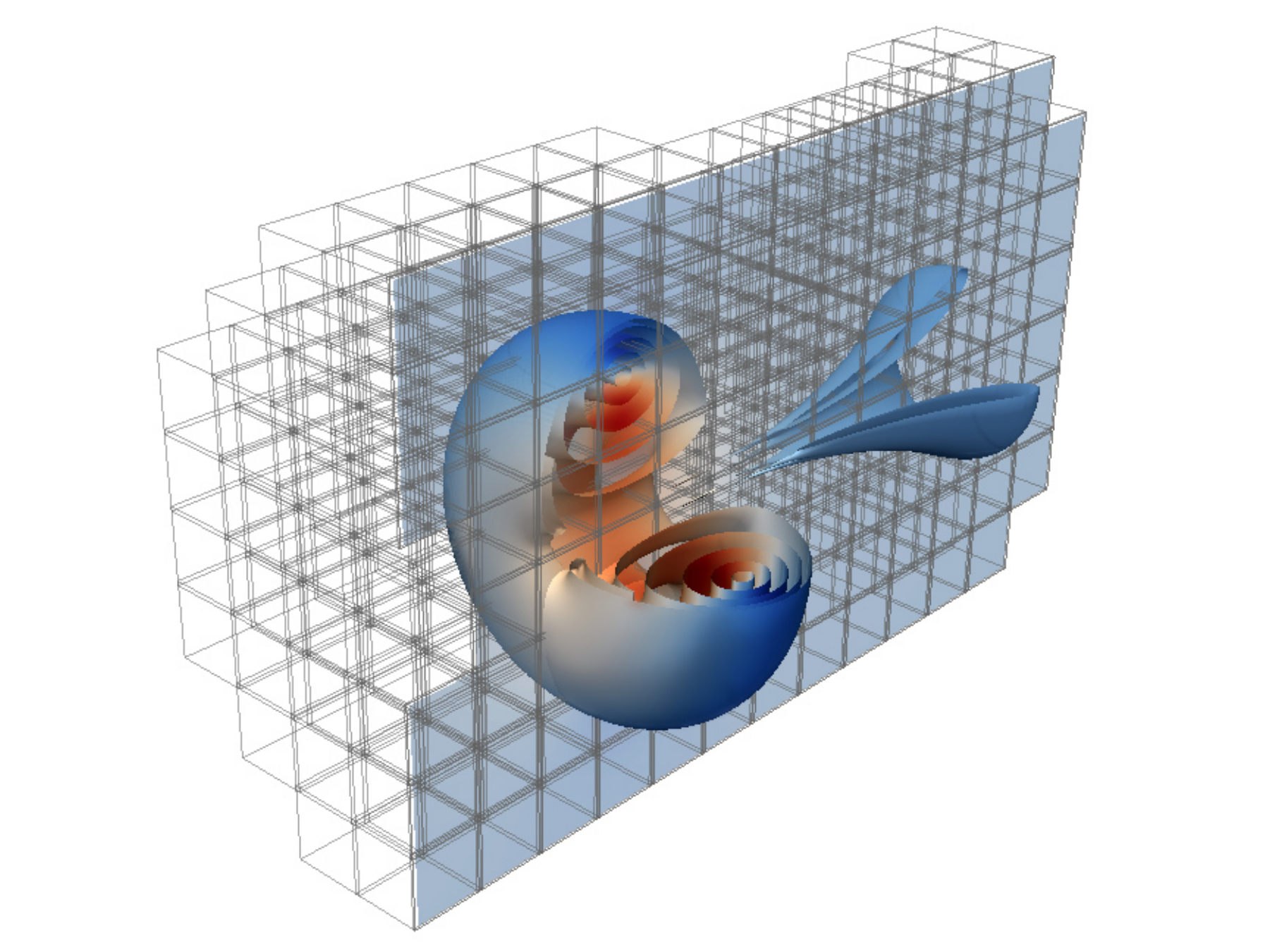}
\caption[%
  Vortex ring at $\text{Re}=\numprint{7500}$ computed using an incompressible Navier-Stokes solver based on the LGF-FMM.
]{%
  Vortex ring at a Reynolds number of $\numprint{7500}$. Isocontours correspond to the absolute value of vorticity (log scale), color corresponds to the streamwise velocity, and gray boxes correspond to the location of grid blocks used in the simulation.\label{fig:vortex_ring}%
}
\end{figure}

The FLGF method is particularly useful for solving PDEs that have been discretized using a numerical scheme that enforces discrete conservation laws.
In such cases, accurate solutions to the difference equations, but not necessarily the original PDE, are necessary to preserve physical fidelity.
We have applied the present method to solve incompressible, viscous, external flows using a finite volume scheme and an infinite staggered Cartesian grid.
Figure~\ref{fig:vortex_ring} includes a snapshot of thin vortex ring at a Reynolds number (based on ring circulation) of 7,500 simulated using this scheme.
A detailed description and results pertaining to the application of the FLGF method to the incompressible Navier-Stokes are the subject of future publications.

In the interest of brevity, our discussion and reported results only pertain to the discrete Laplace kernel, yet the FLGF method can be applied to other non-oscillatory LGFs.
In fact, our method can be readily generalized to any non-oscillatory kernel (including singular kernels); the only restriction is that sources and evaluation points be defined on a regular grid.
Based on these observations, and the simple/standard routines and data-structures involved in the algorithm, it is expected that the FLGF method can be readily incorporated into a wide range of existing methods and codes that solve elliptic PDEs on unbounded domains.

\appendix

\section{Evaluating the LGF of the discrete Laplace operator numerically}
\label{sec:app_lgf}

Values of $\mathsf{G}(\mathbf{n})$ for small $|\mathbf{n}|$ are frequently used by the FLGF method.
Therefore it is advantageous to program an accurate look-up table for values of $\mathbf{n}$ confined to a small cubic box centered at the origin.
The symmetry of $\mathsf{G}(\mathbf{n})$, discussed in Section~\ref{sec:prelim_lgf}, suggests that only approximately $1/48$-th of the total number of points in the box need to be numerically evaluated.
It is possible to reduce the triple integral in Eq.~\ref{eq:lgf1} to a single semi-infinite integral \cite{cserti2000} given by
\begin{equation}
\mathsf{G}(\mathbf{n}) = -\int_{0}^{\infty}
 e^{-6 t}
 I_{n_1}\left( 2 t \right)
 I_{n_2}\left( 2 t \right)
 I_{n_3}\left( 2 t \right)
 \,dt,
\label{eq:lgf3}
\end{equation}
where $I_{k}(x)$ is the modified Bessel function of the first kind of order $k$, and $\mathbf{n}=(n_1,n_2,n_3)\in\mathbb{Z}^3$.
In our experience, it is easier and faster to numerically integrate Eq.~\ref{eq:lgf3} instead of Eq.~\ref{eq:lgf1}.
The integrand of Eq.~\ref{eq:lgf3} is non-oscillatory and smooth throughout the domain of integration.
A simple adaptive Gauss-Kronrod scheme can be used to perform the numerical integration and obtain error estimates.
Furthermore, the semi-infinite integral can be partitioned into two intervals $[0,\alpha]$ and $[\alpha,\infty]$, where $\alpha$ is chosen such that the latter integral can be evaluated analytically using the asymptotic expansion (for large arguments) of $I_n(x)$ \cite{abramowitz1972}.
More efficient implementations might consider partitioning the integration interval into multiple subintervals, exploiting both the ascending series representation and the asymptotic expansion of each Bessel function.

\section{Communication patterns of MPI-based implementation}
\label{sec:app_parallel}

The pseudo-codes provided in this appendix complement the discussion regarding the parallel implementation of the present method included in Section~\ref{sec:flgf_parallel}.
For convenience, in this appendix the term ``node'' is used to denote either grid blocks, grid intervals, or tree-nodes as defined in Section~\ref{sec:flgf_basic} and its precise meaning is deduced from the context.
The algorithms discussed in this appendix are based on non-blocking MPI operations; we refer the reader to \cite{gropp1999} for an introduction to these operations.

\subsection{Level Interactions}
\label{sec:app_parallel_level}

\begin{algorithm}[htbp]
  \setlength{\parskip}{0pt}
 	\While{not done sending or receiving or local-work}{
  	check status of all active messages\;
  	\ForAll{receive-messages that have completed receive}{
   		process receive-message\;
   		mark receive-buffer unit associated with receive-message as available\;
   	}
   	\ForAll{available receive-buffer units}{
   		\If{not done posting receive-messages}{
   			associate receive-message with receive-buffer unit\;
   			post non-blocking receive for receive-message\;
   		}
   	}
   	\ForAll{send-messages that have completed send}{
   		mark send-buffer associated with send-message as available\;
   	}
		\If{not done with all send-work units and send-buffer unit is available}{
			\If{send-work unit corresponds to new send-message}{
   			associate send-message with send-buffer unit\;
   		}
	  	perform $M$ send-work units of send-message\;
  		\If{done building send-massage}{
   			post non-blocking send for send-message\;
   		}
   	}
   	\ElseIf{not done with all local-work units}{
   		perform $N$ local-work units\;
  	}
  }
 \caption[%
  Parallel communication pattern of Level Interactions of the LGF-FMM
  ]{%
  Communication pattern of \emph{Level Interactions}.\label{alg:level}}%
\end{algorithm}

The pseudo-code for the parallel implementation of \emph{Level Interactions}, Step 2 of the FLGF algorithm of Section~\ref{sec:flgf_basic}, is provided in Algorithm~\ref{alg:level}.
A description of the terms and operations of Algorithm~\ref{alg:level} is as follows:
\begin{itemize}[leftmargin=1em]
  \renewcommand\labelitemi{{\boldmath$\cdot$}}
  \setlength{\parskip}{0pt}
	\item \emph{Done sending (receiving)}: all non-blocking MPI messages being sent (received) have been posted and completed.
	\item \emph{Done local-work}: any intra-MPI-process operations have been completed.
	\item \emph{Send-/receive-messages}: As described in Section~\ref{sec:flgf_parallel}, each MPI-process sends, at most, one message to any other MPI-process.
	A message is composed of sub-messages, one for each target node.
	Each sub-message contains information identifying the target node, and the field induced on target node by all source nodes that belong to the sending MPI-process and interact with the target node.
	\item \emph{Buffer and buffer units}: messages are buffered before being posted.
	The send-buffer and receive-buffer are composed of a fixed number of buffer-units.
	Each buffer-unit is allocated enough memory to handle any message that will be posted.
	The examples included in Section~\ref{sec:parallel_tests} use a total of four buffer units, two for sending and two for receiving.
	\item \emph{Process receive-message}: read receive-message and add field contribution from non-local source nodes to local target nodes.
	\item \emph{Send-work unit}: compute the field induced from a single local node to a target node belonging to the current target MPI-process.
	The induced field of each target node is aggregated in send-buffer.
	Operations performed by a single send-work unit correspond to those of a single entry of the sum given in Eq.~\ref{eq:alg_mult}.
	\item \emph{Local-work unit}: same as send-work unit, except that the induced field is added to the local storage of the target node (no need to buffer or perform any MPI communication).
	\item \emph{Done building send-message}: the results from all send-work units required by send-message (or, equivalently, target MPI\hyp{}process) have been packaged into a message.
	\item \emph{Parameters $M$ and $N$}: determine the number of work units performed at each iteration of the main loop.
	The examples included in Section~\ref{sec:parallel_tests} use $M=N=20$.
\end{itemize}

\subsection{Upwards and Downwards Pass}

\begin{algorithm}[htbp]
  \setlength{\parskip}{0pt}
	\ForAll{local nodes}{
		\ForAll{non-local children of node}{
 			post non-blocking receive for message sent by child\;
 		}
 	}
 	\While{not done sending or receiving or local-work}{
  	check status of all active messages and update receive-tracker\;
   	select $M$ ready-for-processing nodes using receive-tracker\;
   	\ForAll{selected nodes}{
			\If{node has children}{
				build node's weights by regularizing child nodes' weights\;
			}
			perform padded-FFT on node's weights\;
			\If{node has a parent}{
				\If{parent is non-local}{
	 				post non-blocking send for message received by parent\;
	 			}
	 			\Else{
	 				update local parent node's weights\;
	 			}
	 		}
	 	}
  }
 \caption{Communication pattern of \emph{Upwards Pass}.\label{alg:up}}%
\end{algorithm}

The pseudo-code for the parallel implementation of \emph{Upwards Pass}, Step 1 of the FLGF algorithm of Section~\ref{sec:flgf_basic}, is provided in Algorithm~\ref{alg:up}.
A description of the terms and operations of Algorithm~\ref{alg:up} is as follows:
\begin{itemize}[leftmargin=1em]
  \renewcommand\labelitemi{{\boldmath$\cdot$}}
  \setlength{\parskip}{0pt}
	\item \emph{Done sending (receiving) and done local-work}: same as in \ref{sec:app_parallel_level}.
	\item \emph{Message}: a messages contain a node's weights (sources or regularized weights/sources from child nodes).
	Each node is provided enough auxiliary storage to receive the weights from all of its children.
	\item \emph{Receive-tracker}: a tree-like data-structure that contains information regarding the progress of all communication and operations associated with each local node.
	It can be used to determine whether a node has finished receiving messages from all of its children (if any), whether a padded-FFT has been performed on its weights, and whether it has posted a non-blocking send to its parent (if any).
	\item \emph{Ready-for-processing node}: a node that has not be processed, but has finished receiving messages from all of its children (if any).
	A node is said to be processed if its weights have been computed (or are known), i.e. Eq.~\ref{eq:alg_reg}, a padded-FFT has been performed on its weights, i.e. Eq.~\ref{eq:alg_upfft}, and a non-blocking send to its parent-node has been posted (if parent exists).
	\item \emph{Selecting ready-for-processing nodes}: the receive-tracker is transversed in leaf-to-root order and nodes that meet the ready-for-processing criteria are selected. Priority is given to nodes whose parent are non-local, i.e. belong to a different MPI-process.
	\item \emph{Parameter $M$}: determine the number of nodes to be processed at each iteration of the main loop.
\end{itemize}
Pseudo-code for \emph{Downwards Pass} is omitted since the communication pattern is very similar that of \emph{Upwards Pass}.
The only significant differences are that nodes send to their children, as opposed to their parent, and that when selecting ready-for-processing nodes the receive-tracker is transversed in root-to-leaf order, as opposed to leaf-to-root order.
The operations performed at each step and the ready-for-processing criteria are readily deduced from the discussion included in Sections~\ref{sec:flgf_basic} and \ref{sec:flgf_parallel}.

\bibliographystyle{model1-num-names}
\bibliography{jcp_flgf}

\end{document}